\documentclass[final, 3p, times,sort&compress]{elsarticle}

\usepackage[colorlinks=true,linkcolor=blue]{hyperref}

\usepackage{amsfonts,amsmath,amssymb,stmaryrd}  
\usepackage{upgreek}                            
\usepackage{graphicx}
\usepackage{siunitx}                            
\usepackage{booktabs}                           
\usepackage[font=footnotesize,skip=0pt,labelfont=bf]{caption}       
\usepackage{nicefrac}

\journal{{\normalfont Accepted manuscript, Composites Science and Technology, \urlstyle{same} \url{https://doi.org/10.1016/j.compscitech.2021.108965}}}

\newcommand{\bs}[1]{\boldsymbol{#1}}
\newcommand{\bt}[1]{\mathbf{#1}}
\newcommand{\tx}[1]{\text{#1}}

\newcommand{\eps}[0]{\varepsilon}
\newcommand{\sig}[0]{\sigma}

\newcommand{\se}[1]{\llbracket #1 \rrbracket}
\newcommand{\av}[1]{\langle #1 \rangle}

\newcommand{\pdif}[2]{\frac{\partial #1}{\partial #2}}

\newcommand{\reff}[1]{Fig. \ref{#1}}
\newcommand{\reft}[1]{Table \ref{#1}}
\newcommand{\refe}[1]{(\ref{#1})}

\usepackage{etoolbox}
\makeatletter
\patchcmd{\ps@pprintTitle}
  {Preprint submitted to}
  {}
  {}{}
\makeatother

\newcommand\blfootnote[1]{%
  \begingroup
  \renewcommand\thefootnote{}\footnote{#1}%
  \addtocounter{footnote}{-1}%
  \endgroup
}

\begin{document}

\begin{frontmatter}

\title{Modeling and simulation of interface failure in metal-composite hybrids}

\author[IFKM]{Franz Hirsch}
\author[IFKM]{Erik Natkowski}
\author[IFKM,DCMS]{Markus Kästner\corref{cor}}
\cortext[cor]{Corresponding author}
\ead{markus.kaestner@tu-dresden.de}

\address[IFKM]{Institute of Solid Mechanics, TU Dresden, 01062 Dresden, Germany}
\address[DCMS]{Dresden Center for Computational Materials Science (DCMS), TU Dresden, 01062 Dresden, Germany}

\blfootnote{© 2021. This manuscript version is made available under the CC-BY-NC-ND 4.0 license}
\blfootnote{https://creativecommons.org/licenses/by-nc-nd/4.0/}

\begin{abstract}
The application of hybrid composites in lightweight engineering enables the combination of material-specific advantages of fiber-reinforced polymers and classical metals. The interface between the connected materials is of particular interest since failure often initializes in the bonding zone. In this contribution the connection of an aluminum component and a glass fiber-reinforced epoxy is considered on the microscale. The constitutive modeling accounts for adhesive failure of the local interfaces and cohesive failure of the bulk material. Interface failure is represented by cohesive zone models, while the behavior of the polymer is described by an elastic-plastic damage model. A gradient-enhanced formulation is applied to avoid the well-known mesh dependency of local continuum damage models. The application of numerical homogenization schemes allows for the prediction of effective traction-separation relations. Therefore, the influence of the local interface strength and geometry of random rough interfaces on the macroscopical properties is investigated in a numerical study. There is a positive effect of an increased roughness on the effective joint behavior.
\end{abstract}

\begin{keyword}
Hybrid composites\sep Matrix cracking\sep Interface \sep Damage mechanics \sep Material modelling \sep Cohesive failure \sep Gradient-damage
\end{keyword}

\end{frontmatter}

\section{Introduction}
\begin{figure}
 \centering
 \linespread{1.3}
 \includegraphics{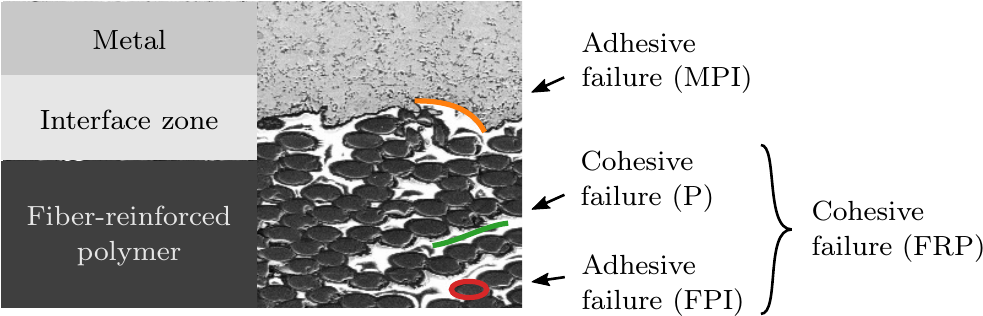}
 \caption{Microsection of a random rough interface between metal and fiber-reinforced polymer. Adhesive failure of the metal-polymer interface (MPI) occurs alongside cohesive failure of the fiber-reinforced polymer (FRP). The latter is a summary of adhesive failure of the fiber-polymer interface (FPI) and cohesive failure of the polymer (P).}
\label{fig:motivation}
\end{figure}

The connection of different materials in hybrid structures needs special attention, since the failure of joints can lead to a complete collapse of engineering constructions. Conventional concepts like screw connections \cite{pramanik2017} or rivet joints \cite{hirsch2017a,rao2020} often induce undesired pre-damage within the composite, while the integration of additional adhesives increases the complexity of the manufacturing process \cite{zinn2018}. Interlocking non-destructive approaches represent a suitable alternative \cite{barfuss2018,nieschlag2019,herwig2018}, where the connection is created during a hybridization step. All mentioned approaches exhibit load-bearing interfaces, where failure typically initializes. One reason for this is the low interfacial strength compared to the cohesive strength of the connected materials \cite{yao2002,zhao2007}.  
\\
The microscopic interface roughness has a significant influence on the effective interface strength and can improve the mechanical properties of the metal-composite joint considered in this contribution. There exist several experimental and numerical studies concerning the influence of structured interfaces on the overall joint properties. An increased roughness can be achieved with e.g. a laser pre-treatment or sandblasting of the metal surface \cite{kiessling2017}. The choice of pre-treatment  depends on the desired profile, like randomly distributed height profiles \cite{lucchetta2011} or defined surface patterns \cite{kim2010,cordisco2016,hosseini2019}. This enables the polymer to fill the gaps during the manufacturing process and to form a microscopic contour joint.  The enhancement of the local surface roughness can increase the effective strength, e.g. due to an extended crack path along the interface \cite{cordisco2016} or a transition from local adhesive to cohesive failure \cite{kim2010}. 
\\
The understanding of damage phenomena is valuable during the design process and often hard to obtain experimentally. Modeling and simulation techniques are used more  frequently in such situations and enable the identification of individual structure-property relationships. Cohesive zones \cite{barenblatt1962} represent a common modeling approach to describe interface failure. The effect of an increased roughness can be considered in this case, e.g., in a smeared way with a flat geometry and increased cohesive zone properties. In contrast, there are models within a fracture mechanics concept that explicitly resolve the local interface. This makes it possible to investigate the amount of adhesive and cohesive failure depending on the interface roughness \cite{yao2002,noijen2009}. A combination of cohesive zone models with a sinusoidal interface for similar \cite{Zavattieri2007,Zavattieri2008} and dissimilar materials \cite{Cordisco2012,Cordisco2014} was also presented for studies of pure adhesive failure. Recently, Hirsch and Kästner \cite{hirsch2017,hirsch2018,hirsch2018a} proposed a modeling strategy, where adhesive and cohesive failure of a bi-material interface is considered within a computational homogenization scheme. A cohesive zone model is used to describe adhesive failure of the local interface and continuum damage mechanics to incorporate cohesive failure of the bulk material. Therefore, effective properties of geometrically idealized interfaces can be determined based on representative volume elements. 
\\
There are several homogenization schemes in the literature for thin inhomogeneous layers, where an effective traction-separation law is obtained instead of a stress-strain relation \cite{Matous2008,cidalfaro2010,Verhoosel2010}. While the referred contributions are formulated in a small deformation framework, Hirschberger et al. \cite{hirschberger2009} extend the approach of Matou\v{s} et al. \cite{Matous2008} to finite deformations.
\\
The current paper enhances the procedure of Hirsch and Kästner \cite{hirsch2017} to the general case of a fiber-reinforced polymer (FRP) connected to a metal component in the context of large deformations. The choice of the representative interface element is motivated by the microsection shown in \reff{fig:motivation}. The metal material has a random rough surface, so that the fibers can interact with it and the polymer can fill the gaps. Under loading, different inelastic phenomena occur until complete failure: adhesive failure of the metal-polymer interface and cohesive failure of the FRP. Within this contribution, the latter has to be understood as cohesive failure of the homogenized FRP. Thus, it is a combination of cohesive failure of the polymer and adhesive failure of the fiber-polymer interface as well. The adopted homogenization step is based on the approach of Hirschberger et al. \cite{hirschberger2009} and is applied to a boundary layer of a predefined height. Failure of the microstructure is considered with cohesive zone models for all material interfaces and a continuum damage model for the polymer material. A gradient formulation based on the framework of Peerlings et al. \cite{peerlings1996,peerlings2001} is used to reduce the well-known mesh dependency of local continuum damage models.
\\
The paper is structured as follows: the constitutive models including the employed constitutive interface description are provided in Section \ref{sec:constitutive _models}. The numerical model of the microstructure is described in Section \ref{sec:computational_framework}. It consists of a brief introduction of the applied homogenization approach, the geometry generation and the parametrization of all  models. Finally, the failure behavior of several microstructures is studied. Results are presented and discussed in Section \ref{sec:results}.

\section{Constitutive models}
\label{sec:constitutive _models}
The constitutive equations for the different materials of the hybrid joint include models for the metal, fiber and polymer materials as well as their interfaces. Within the concept of continuum mechanics, a body takes the reference configuration $\mathcal B_0 \subset \mathcal R^3$ at time $t_0$ with the domain boundary $\partial\mathcal B_0$ and the current configuration $\mathcal B \subset \mathcal R^3$ at time $t > t_0$ with the domain boundary $\partial\mathcal B$. Material points are identified by their position vectors $\bt X \in \mathcal B_0$ and $\bt x =\bs\varphi(\bt X,t) \in \mathcal B$, where $\bs\varphi$ describes the motion of the material body. Their displacement is characterized by the vector $\bt u(\bt X,t)=\bs\varphi(\bt X,t)-\bt X$. The deformation gradient and its determinant are defined as
\begin{equation}
 \bt F=\pdif{\bs\varphi}{\bt X} ~\tx{and}~ J=\tx{det}(\bt F) > 0.
\end{equation}
Furthermore, the left Cauchy-Green deformation tensor $\bt b = \bt F \cdot \bt F^\tx{T}$ and the velocity gradient $\bt l=\partial\bt v/\partial \bt x$ are introduced with  $\bt v=\dot{\bt u}$ being the velocity.

\subsection{Metal and fiber material}
The metal and fiber materials, i.e., the aluminum and glass fiber components are assumed to be isotropic and elastic. Hence, the material response is determined by the free energy $\Psi(\bt b)=\Psi(\lambda_1,\lambda_2,\lambda_3)$, where $\lambda_i$ are the principal stretches. A dissipation rate $\mathcal D$ per unit of reference volume is formally introduced for the purely mechanical case to satisfy the Clausius-Duhem inequality (CDI)
\begin{equation}
\label{eq:dissipation_1}
\mathcal D=\bs\tau:\bt d - \dot\Psi(\bt b) \geq 0,
\end{equation}
with the Kirchoff stress tensor $\bs \tau$ and the deformation rate $\bt d=1/2 \left( \bt l + \bt l^\tx T \right)$. Evaluation of \refe{eq:dissipation_1} using the procedure of Coleman and Noll \cite{coleman1963} yields the stress relation 
\begin{equation}
\label{eq:stress_deformation}
 \bs\tau=2\pdif{\Psi}{\bt b}\cdot \bt b. 
\end{equation}
The Cauchy stress tensor $\bs\sigma$ can be determined with $\bs \sigma = 1/J\bs\tau$. Here, the general phenomenological ansatz  
\begin{equation}
\label{eq:ogden}
 \Psi=\sum\limits_{p=1}^{N}\frac{\mu_p}{\alpha_p} \left( \check \lambda_1^{\alpha_p}+\check \lambda_2^{\alpha_p} + \check \lambda_3^{\alpha_p} -3 \right) + \frac{\kappa}{4}\left( J^2-2\ln J - 1 \right)
\end{equation}
of Ogden type~\cite{ogden1972} is employed for both materials. While $\mu_p, \alpha_p, \kappa$ represent material parameters, $\check \lambda_i=\lambda_i J^{-1/3}$ are the isochoric principal stretches according to the  Flory-split~\cite{flory1961} and an additive split of $\Psi$ in a deviatoric and volumetric part in \refe{eq:ogden}.

\subsection{Matrix material}
Many thermoplastics and thermosets used in composites show a complex material response depending on the material state and the environmental conditions. For the considered epoxy system, a stress-state-dependent yield and fracture behavior can be observed in uniaxial tension and compression tests \cite{fiedler2001}. The objective of constitutive modeling in this section is to capture some key effects such as the strongly nonlinear stress-strain response and the pressure sensitive failure. Hence, other influencing factors like loading rate, temperature and aging are not considered in this work.
\\
The formulation of the constitutive equations is based on classical continuum damage mechanics. Within the concept of effective stresses \cite{lemaitre1985,lemaitre1996}, the nominal Kirchhoff stress $\bs\tau$ of a damaged continuum is related to the effective Kirchhoff stress $\bs{\hat\tau}$ of an equivalent undamaged continuum in terms of
\begin{equation}
\label{eq:tau_damage}
 \bs\tau=(1-D)\bs{\hat\tau},
\end{equation}
with $D \in  \mathcal R_+ $ and $ D\leq 1$ being a non-decreasing scalar damage variable. The undamaged material state corresponds to $D=0$ and the fully damaged state with no further stress transmission to $D=1$. The undamaged stress response and evolution equations follow the large strain elastoplastic model presented by Simo~\cite{simo1992}, where a multiplicative split of the deformation gradient
\begin{equation}
 \bt F = \bt F^\tx e \cdot \bt F^\tx p
\end{equation}
into an elastic part $\bt F^\tx e$ and a plastic part $\bt F^\tx p$ is employed \cite{lee1969}. The corresponding dissipation function reads
\begin{equation}
\label{eq:dissipation}
\mathcal D=\bs\tau:\bt d - \dot\Psi(\bt b^{\tx e}) \geq 0,
\end{equation}
with the 'damaged' free energy function $\Psi=(1-D)\hat\Psi$ and the elastic left Cauchy-Green deformation tensor $\bt b^\tx e = \bt F^\tx e \cdot \bt F^\tx{eT}$. The evaluation of \refe{eq:dissipation}, in accordance to \cite{coleman1963}, leads with the Lie derivative $\mathcal L_v \bt b^\tx e$ to the stress relation and the reduced dissipation inequality
\begin{equation}
\label{eq:dissipation_reduced}
 \bs\tau=2(1-D)\pdif{\hat\Psi}{\bt b^\tx e}\cdot \bt b^\tx e 
 \quad\tx{and}\quad 
 \bs\tau : \left[ -\frac{1}{2}\left( \mathcal L_v \bt b^\tx e \right) \cdot\left(\bt b^\tx{e}\right)^{-1} \right] + \hat\Psi\dot D \geq 0.
\end{equation}
To satisfy the reduced dissipation inequality \refe{eq:dissipation_reduced}, suitable evolution equations for $\bt b^\tx e$ and $D$ have to be defined.

\subsubsection*{Elasto-plastic response}
The elastic response of the model is determined by the 'undamaged' free energy $\hat\Psi(\bt b^\tx e)=\hat\Psi(\lambda^{\tx e}_1,\lambda^{\tx e}_2,\lambda^{\tx e}_3)$, where $\lambda^{\tx e}_i$ are the principal stretches of $\bt b^\tx e$. For $\hat \Psi$ also the proposed Odgen type Ansatz in \refe{eq:ogden} is employed. 
Plastic flow occurs, if a yield condition is fulfilled. For polymers, often a parabolic yield criterion is used to describe a pressure sensitive or an asymmetric tension-compression behavior. Therefore, the parabolic criterion proposed by Tschoegl~\cite{tschoegl1971}
\begin{equation}
\label{eq:yield}
 f=3J_2 + (\sig_\tx c-\sig_\tx t)I_1 - \sig_\tx t \sig_\tx c
\end{equation}
is applied. In equation \refe{eq:yield}, $J_2=1/2~\bt{\hat s}:\bt{\hat s}$ represents the second invariant of the effective Cauchy stress deviator $\bt{\hat s}=\bs{\hat \sigma}-1/3~I_1 \bt I$ in which $\bs{\hat\sigma}=1/J~\hat{\bs\tau}$ and $I_1=\tx{tr}\left( \bs{\hat\sigma} \right)$ are the effective Cauchy stress tensor as well its first invariant and  $\bt I$ the identity tensor. Furthermore, isotropic hardening is determined by the tensile and compressive yield strengths $\sigma_\tx t$ and $\sigma_\tx c$.

The evolution of the elastic left Cauchy-Green deformation tensor is governed by the non-associative flow rule
\begin{equation}
\label{eq:flow_rule}
 -\frac{1}{2}\left( \mathcal L_v \bt b^\tx e \right) \cdot\left(\bt b^\tx{e}\right)^{-1} = \dot \lambda \pdif{Q}{\bs{\hat\sigma}},
\end{equation}
with the plastic multiplier $\dot\lambda$ according to the loading-unloading conditions $\dot\lambda \geq 0,~f \leq 0,~\dot\lambda f = 0$ and the consistency condition $\dot f =0$. The plastic potential
\begin{equation}
\label{eq:plastic_potential}
Q=3 J_2+\alpha I^2_1
\end{equation}
is defined with respect to the effective Cauchy stress tensor $\bs{\hat\sigma}$ and enables the control of plastic Poisson effects by the parameter $\alpha$. The relation $\alpha=(9-18\nu_\tx p)/(2+2\nu_\tx p)$ with respect to the initial plastic Poisson ratio $\nu_\tx p$ as presented by \cite{melro2013} is adopted. The derivative in the flow rule \refe{eq:flow_rule} could be defined alternatively with the nominal stress tensor but is left with the (undamaged) effective stress tensor. This is due to advantages in the numerical implementation, similar to the model proposed by Mediavilla et al.~\cite{mediavilla2006}. In contrast to the flow rule proposed by Simo~\cite{simo1992} -- where the Kirchhoff stress tensor is utilized -- \refe{eq:yield}-\refe{eq:plastic_potential} are formulated with the Cauchy stress tensor, since yield stresses and strength values are often evaluated based on Cauchy or first Piola-Kirchhoff stresses~\cite{meschke1999}. 

The tensile and compressive yield strengths $\sig_\tx t$ and $\sig_\tx c$ in \refe{eq:yield} are defined by the evolution equations
\begin{equation}
\label{eq:harding_evo}
    \begin{aligned}
        \dot\sig_\tx t &= h_\tx t(\eps^\tx{p}) \dot \eps^\tx{p} \quad\tx{with}& \sig_\tx t(0) &= \sig_\tx{t}^0 \quad\tx{and} \\
        \dot\sig_\tx c &= h_\tx c(\eps^\tx{p}) \dot \eps^\tx{p} \quad\tx{with}& \sig_\tx c(0) &= \sig_\tx{c}^0.
    \end{aligned}
\end{equation}
The initial yield stresses $\sig_\tx{t}^0$ and $\sig_\tx{c}^0$ and the functions for the hardening moduli $h_\tx t$ and $h_\tx c$ must be chosen to describe the experimental results of tension and compression tests phenomenologically. Isotropic hardening is governed by the equivalent plastic strain and defined by \cite{melro2013} 
\begin{equation}
 \dot \eps^\tx{p}=k\dot\lambda\sqrt{\pdif{Q}{\bs{\hat\sigma}}:\pdif{Q}{\bs{\hat\sigma}}},
\end{equation}
where $k=1/(1+2\nu^2_\tx p)^{1/2}$ depends on the initial plastic Poission ratio $\nu_\tx p$.

\subsubsection*{Gradient-enhanced damage evolution}
Similiar to the yield condition in \refe{eq:yield}, a pressure sensitive damage activation function is introduced to control damage initiation. The damage activation function 
\begin{equation}
\label{eq:damage_activation}
 F=\frac{3}{X_\tx t X_\tx c} J_2 + \frac{X_\tx c - X_\tx t}{X_\tx t X_\tx c}  I_1 -r \leq 0 
\end{equation}
proposed by Melro et al. \cite{melro2013} is adopted, where $X_\tx t$ and $X_\tx c$ represent the tensile and compressive strength of the material. The internal variable $r$ controls the evolution of $F$. With another set of Kuhn-Tucker conditions $\dot r \geq 0, F \leq 0$ and $\dot r F =0$, damage loading and unloading situations can be distinguished. The evolution of the damage variable $D$ is governed by
\begin{equation}
\label{eq:damage_evo}
 \dot D=h_\tx D(\kappa,D)\dot{\kappa},
\end{equation}
where $\kappa$ is a damage driving internal variable. 
\\
The application of classical damage models leads to localization phenomena and FE mesh dependent results. To overcome these problems, different regularization methods exist. In the current model, a gradient enhancement based on the framework of Peerlings et al. \cite{peerlings1996,peerlings2001} is employed. To this end, the damage driving variable $\kappa$ is defined as a non-decreasing non-local variable
\begin{equation}
 \kappa =  \underset{\tau \leq t}{\tx{max}}\left[\bar\eps(\tau)\right].
\end{equation}
The non-local variable $\bar \eps$ is connected to a local counterpart $\tilde\eps$ by a Helmholz-type equation
\begin{equation}
 \bar\eps-l^2\Delta\bar\eps=\tilde\eps,
\end{equation}
which forms an additional boundary value problem. A homogeneous Neumann boundary condition $\nabla\bar\eps \cdot \bt n =0$ is assumed using the outward normal vector $\bt n$ on $\partial\mathcal B$. Both, the Laplacian and the Nabla operator $\Delta$ and $\nabla$, are defined with respect to the current configuration. The length parameter $l$ ensures a damage localization within a defined volume instead of a single element to avoid the spurious mesh dependency. If the failure condition in \refe{eq:damage_activation} is fulfilled, the local damage driving variable
\begin{equation}
 \dot{\tilde\eps} = \left\{\begin{array}{ll} 
                        \dot\eps^\tx p & \tx{for } \dot r > 0\\
                        0  & \tx{for }\dot r = 0
                    \end{array}\right. 
\end{equation}
evolves simultaneously to the equivalent plastic strain. All the proposed models were implemented in the commercial software package ABAQUS \cite{dassaultsystemes2018} as user defined subroutines according to the strategy presented by Seupel at al. \cite{seupel2018}. 

\subsection{Interfaces}
The constitutive law for the metal-polymer and the fiber-polymer interfaces is based on the cohesive zone model provided by ABAQUS \cite{dassaultsystemes2018}. Therefore, a simple uncoupled traction-separation law (TSL) for interfaces with zero thickness is used. Damage initiation is determined by the quadratic stress criterion
\begin{equation}
 \left( \frac{\langle t_\tx n \rangle_{\scriptscriptstyle +}}{t^0_\tx n} \right)^2 + \left( \frac{ t_{\tx s_1}}{t^0_{\tx s_1}} \right)^2 + \left( \frac{ t_{\tx s_2}}{t^0_{\tx s_2}} \right)^2 -1=0,
\end{equation}
where the traction coordinates $t_I$ are compared to the corresponding pure mode strengths $t^0_I$. In case of a pure state of compression no damage is initiated by using of the Macaulay brackets $\langle \cdot \rangle_{\scriptscriptstyle \pm}=1/2 (\cdot \pm |\cdot|)$. After a linear damage evolution, the fully damaged state is governed by the fracture criterion
\begin{equation}
 \left( \frac{G_\tx n}{G^\tx C_\tx{n}} \right)^2 + \left( \frac{G_{\tx{s}_1}}{G^\tx C_{\tx{s}_1}} \right)^2 + \left( \frac{G_{\tx{s}_2}}{G^\tx C_{\tx{s}_2}} \right)^2 -1=0,
\end{equation}
with $G_I$ indicating the work done by the tractions and $G^\tx C_I$ being the critical fracture energies in the normal and the two shear directions.

\section{Computational framework}
\label{sec:computational_framework}
\begin{figure}
 \centering
 \linespread{1.3}
 \includegraphics{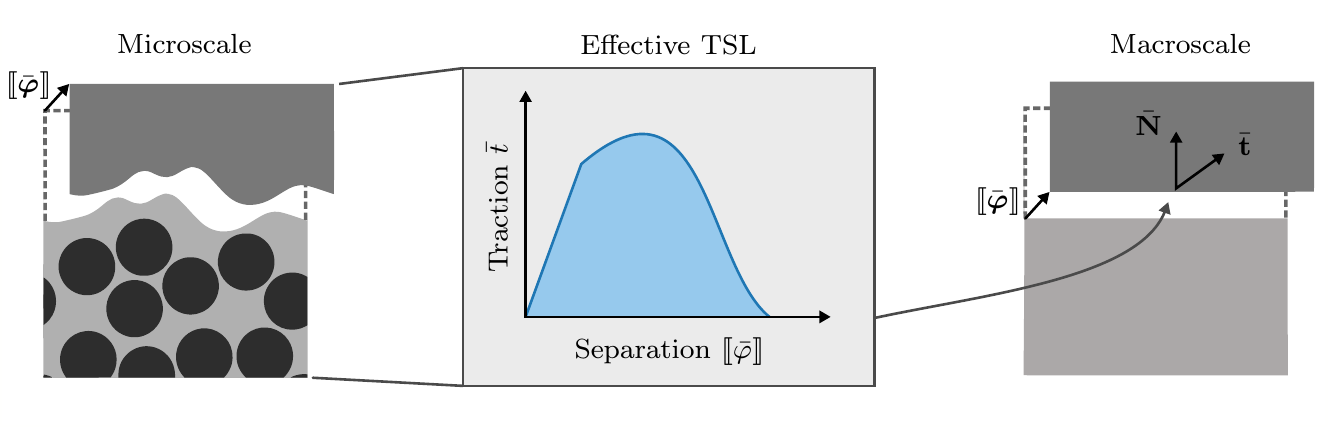}
 \caption{Multiscale scheme for interface regions.}
\label{fig:homogenisation_scheme}
\end{figure}

This section provides the general framework for the analysis of interface regions between FRP and metal components as shown in \reff{fig:homogenisation_scheme}. A respresentative volume element (RVE) with a heterogenous microstructure, i.e. in matrix material embedded fibers and a random rough interface, is investigated. Effective traction-separation relations result from a homogenization and enable the extraction of effective interface properties, which can be used further for a macroscopically flat interface.

\subsection{Scale transition}
\begin{figure}
 \centering
 \footnotesize
 \linespread{1.3}
 \includegraphics{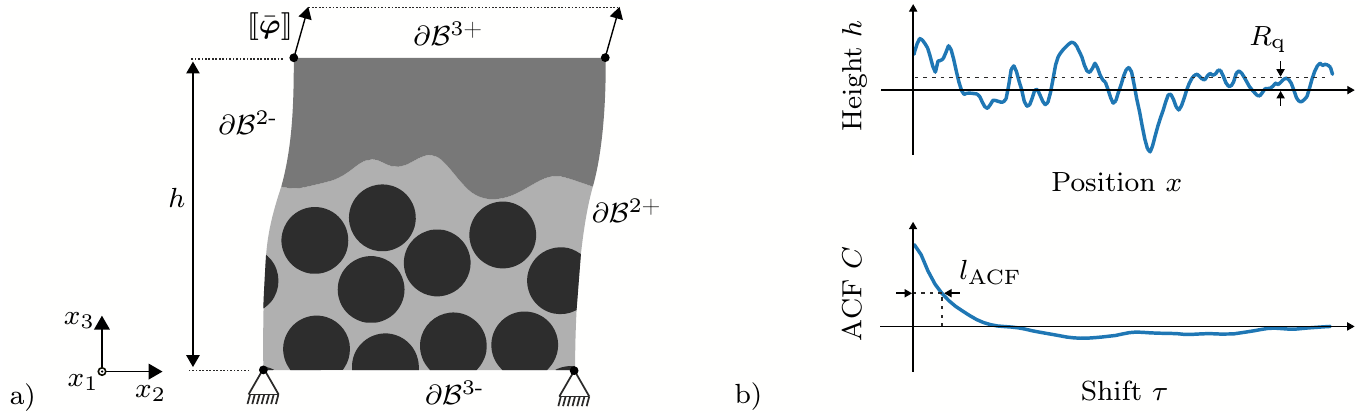}
 \caption{a) Deformed RVE configuration with hybrid boundary conditions and b) characteristic measures of height profile $h(x)$ with the root-mean-square roughness $R_\tx q$ and the correlation length $l_\tx c$ of the autocorrelation function $C$.}
\label{fig:hybrid_boundary_conditions_rq_lc}
\end{figure}

The effective response on the macroscale is determined by the volume averages of microscopic field quantities. i.e. $\av{\bullet} = 1/V_0\int_{\mathcal B_0} \bullet ~\tx dV$. A physically meaningful connection between both scales is achieved by the Hill-Mandel condition \cite{hill1963} and reads
\begin{equation}
\label{eq:hill_general}
 \av{\bt P} : \av{\dot{\bt F}} = \av{\bt P : \dot{\bt F}},
\end{equation}
with the first Piola-Kirchhoff stress tensor $\bt P=\bt F^{-1} \cdot \bs \tau$. Different derivations of this condition exist for thin layers and interfaces, where macroscopic traction-separation relations are provided instead of stress-strain relations. For large deformations the framework proposed by Hirschberger et al. \cite{hirschberger2009} is employed. It relates the macroscopic interface kinematics $\bar{\bs \varphi}$ and interface traction $\bar{\bt t}$ with
\begin{equation}
\label{eq:effective_tsl}
    \av{\bt F}  = \bt I + \frac{1}{h_0}\se{\bar{\bs\varphi}} \otimes \bar{\bt N} \qquad \tx{and} \qquad 
    \bar{\bt t} = \bar{\bt N} \cdot \av{\bt P} 
\end{equation}
to the volume averages of the microscopic deformation gradient $\av{\bt F}$ and the first Piola-Kirchhoff stress $\av{\bt P}$. In equation \refe{eq:effective_tsl}, $h_0$ represents the initial RVE height, $\overline{\bt N}$ the initial unit normal vector  of a macroscopic interface and $\se{\cdot}$ the jump operator. Inserting \refe{eq:effective_tsl} into \refe{eq:hill_general} leads to an equivalent condition for interfaces
\begin{equation}
\label{eq:hill_interface}
 \bar{\bt t} \cdot \se{\dot{\bar{\bs\varphi}}} = h_0 \av{\bt P : \dot{\bt F}}.
\end{equation}
Appropriate boundary conditions have to be defined to fulfill condition \refe{eq:hill_interface}. Hybrid boundary conditions are suitable for interface layers \cite{hirschberger2009} and are adopted in this work, see \reff{fig:hybrid_boundary_conditions_rq_lc}~a). These boundary conditions consist of a combination of prescribed displacements on the bottom and top face of the RVE and periodic displacements with antiperiodic fluxes on the remaining faces:
\begin{align}
\label{eq:pbc_u}
 \mathcal U &= \left\{ \bt u \in \mathcal R^3 ~|~ \bt u = \left( \bar{\bt F} - \bt I \right) \cdot \bt X +\tilde{\bt u} \tx{ with }\tilde{\bt u}^{1,2+}=\tilde{\bt u}^{1,2-} \tx{ and } \tilde{\bt u}^{3+}=\tilde{\bt u}^{3-}=0  \right\} \tx{ and } \\
 \label{eq:pbc_t}
 \mathcal N&=\left\{ \bt t \in \mathcal R^3 ~|~ \bt t = \bt N \cdot \bt P \tx{ with } \bt{t}^{1,2+}=-\bt{t}^{1,2-} \right\}
\end{align}
Here, $(\cdot)^{\alpha\pm}$ denote the associated values on opposite RVE boundaries $\partial\mathcal B^{\alpha\pm}_0$ in direction $\alpha\in\{1,2,3\}$ with the outward normal vector $\bt N$ and $\tilde{(\cdot)}$ respresents fluctuation quantities. A discussion of suitable boundary conditions of non-local quantities within an RVE framework can be found in \cite{nguyen2019}. This approach is adopted in the sense of hybrid boundary conditions. For the non-local field quantity $\bar\eps$ homogeneous Neumann boundary conditions are used for all (inner) material boundaries while periodic boundary conditions in accordance to the periodic displacement field, i.e. 
\begin{align}
 \mathcal E &=\left\{\bar\eps \in \mathcal R ~|~ \bar\eps^{1,2+} =  \bar\eps^{1,2-} \right\} \tx{ and} \\ 
 \mathcal M &=\left\{ p \in \mathcal R ~|~ p = J \bt F^{-1}\cdot \nabla\bar\eps \cdot \bt N \tx{ with }  p^{1,2+} = -p^{1,2-} \tx{ and } p^{3-}=p^{3+}=0  \right\}
\end{align}
are applied on the RVE boundary.

The hybrid boundary conditions can further be used to transform the effective traction in \refe{eq:effective_tsl}. The incorporation of the volume average of the microscopic first Piola-Kirchhoff stress
\begin{equation}
 \av{\bt P}=\frac{1}{V_0}\int\limits_{\mathcal B_0} \bt P \tx dV = \frac{1}{V_0}\int\limits_{\partial\mathcal B_0}\bt X \otimes \bt t \,\tx dA
\end{equation}
and the antiperiodic traction \refe{eq:pbc_t} in the definition of the effective traction \refe{eq:effective_tsl} results in
\begin{equation}
\label{eq:effective_tractions}
 \bar{\bt t} = \frac{1}{A_0^{3+}}\int\limits_{\partial\mathcal B^{3+}_0} \bt t \,\tx dA,
\end{equation}
with the area $A_0^{3+}$ of the positive boundary $\partial\mathcal B^{3+}_0$ of the undeformed RVE. Hence, the computation of the effective traction for arbitrary microstructures is performed with the average of the microscopic traction over the top surface of the RVE. In \refe{eq:effective_tractions} and accordingly in \refe{eq:effective_tsl} it is assumed, that the unit normal vector $\overline{\bt N}$ of the macroscopic interface is aligned with the local 3-direction.

\subsection{RVE generation}
\label{sec:rve_generation}
The RVE generation consists of two main steps and is implemented within a Python script for an automatic model generation in ABAQUS. In the first step the interface has to be generated. Therefor, a 'random field surface generator' was implemented based on the algorithm proposed by Temizer et. al \cite{temizer2011}. After that, fibers are placed with a modified version of the 'random distribution algorithm' proposed by Melro et. al \cite{melro2008}. For the surface generation, a uniform random generator is used to create a field of random numbers in a pixel space. Afterwards, a Gaussian filter is applied multiple times to smooth the random field. The resulting discrete surface is then translated to a NURBS description \cite{penguian2020} and exported as a CAD geometry for ABAQUS. In this algorithm, geometric periodicity is guaranteed for an easy generation of periodic FE meshes and node-wise application of periodic boundary conditions. Two filter loops are performed with the filter parameters $r=a=5$, see \cite{temizer2011}, equations (2.1) and (2.2). There exist several roughness parameters to describe a surface, where the specific choice depends, among other things, on the individual application. Common amplitude parameters are the mean roughness $R_\tx a$, the root-mean-square roughness $R_\tx q$ or the average maximum height of the profile $R_\tx z$, while the mean peak width $RS_\tx m$ is an example of a common spacing parameter. Here, the root-mean-square (RMS) roughness
\begin{equation}
\label{eq:rms}
 R_\tx q=\sqrt{\frac{1}{L}\int\limits_{0}^{L} \left( h(x) - \overline h  \right)^2 \tx dx} \qquad\tx{with}\qquad \overline h  = \frac{1}{L}\int\limits_{0}^{L} h(x) \,\tx dx
\end{equation}
is used as a characteristic measure for the height distribution of the surface. In \refe{eq:rms}, $h(x)$ denotes the height profile over the coordinate $x$ and $L$ is the sampling length as shown in \reff{fig:hybrid_boundary_conditions_rq_lc}~b). Furthermore, the profile spacing is evaluated by the correlation length $l_\tx{ACF}$. This measure was already used in literature for polymer-metal interfaces, e.g. by Yao and Qu \cite{yao2002}. The normalized autocorrelation function (ACF)
\begin{equation}
 C(\tau)=\lim_{L\rightarrow\infty} \frac{1}{\sig^2L}\int\limits_0^L h(x)h(x+\tau)\,\tx dx,
\end{equation}
with the standard deviation $\sig$ and shift $\tau$, see \reff{fig:hybrid_boundary_conditions_rq_lc}~b), gives the correlation of a height profile with a delayed copy of itself. With zero shift, the height profile is completely correlated and the ACF reaches the value 1. With increasing delay, the correlation decreases. The correlation length $l_\tx{ACF}$ is used as a characteristic spacing parameter and is defined as the shift, where the ACF reaches a certain threshold $C(\tau=l_\tx{ACF})=C_\tx{thres}$. As proposed in \cite{yao2002}, a threshold of $C_\tx{thres}=0.5$ is used to calculate $l_\tx{ACF}$ of a specific surface. Finally, a dimensionless roughness parameter 
\begin{equation}
\label{eq:roughness}
 R=\frac{R_\tx q}{l_\tx{ACF}}
\end{equation}
is introduced as the ratio of the amplitude parameter $R_\tx q$ and the spacing parameter $l_\tx{ACF}$. After the surface generation, fibers with a radius of $r_\tx F=l_\tx{ACF}$ are placed in the section of matrix material within the RVE. Finally, the assembly, the mesh generation and application of periodic boundary conditions is performed script controlled with Python in ABAQUS .

\subsection{Parameter identification}
\label{sec:parametrization}
\begin{figure}
 \centering
 \linespread{1.3}
 \includegraphics{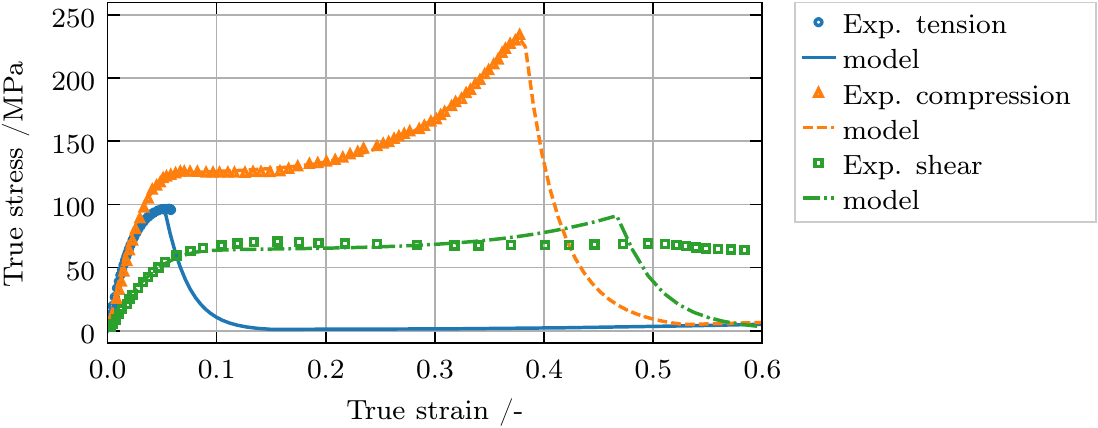}
 \caption{Comparison of the applied hyperelastic-plastic gradient damage model with experimental test data of an epoxy resin taken from \cite{fiedler2001}.}
\label{fig:fitting}
\end{figure}

In this section, the parameter identification of the constitutive models for the fibers, the matrix, the metal component and the metal-matrix as well as fiber-matrix interfaces is addressed. The fibers should mimic E-glass fibers and are parametrized with typical elastic constants taken from \cite{soden1998,melro2013a,schuermann2007}. 
Using $N=1$, $\alpha_1=2, \mu_1=0.5~E/(1+\nu)$ and $\kappa=1/3~E/(1-2\nu)$, they are incorporated into the free energy \refe{eq:ogden} and yield a neo-Hookean model.
The metal phase is parametrized with common elastic constants of a structural aluminum alloy \cite{davies1994}. The parameters for both models are summarized in \reft{table:elastic_constants}.
\begin{table}
\begin{minipage}[t]{0.4\textwidth}
    \centering
    \linespread{1.3}
    \caption{Mechanical material parameters\strut}
    \label{table:elastic_constants}
    \begin{tabular}{l*{3}{r}}     \toprule
     material      & $\nicefrac{E}{\si{\giga\pascal}}$& $\nu $& $\nu_\tx p$  \\ \midrule
     glass fiber   & 74   & 0.2  & -  \\
     aluminum      & 70   & 0.34 & - \\
     epoxy         & 3.75 & 0.39 & 0.3 \\ \bottomrule
    \end{tabular}
\end{minipage}
\begin{minipage}[t]{0.6\textwidth}
    \centering
    \linespread{1.3}
    \caption{Parameters of the hardening model\strut}
    \label{table:hardening}
    \begin{tabular}{*{6}{c}}     \toprule
       $\nicefrac{\sig^0_\tx t}{\si{\mega\pascal}}$& $\nicefrac{a_1}{\si{\mega\pascal}}$ & $a_2$ & $\nicefrac{a_3}{\si{\mega\pascal}}$ & $a_4$ & $\nicefrac{\sig_\tx t}{\sig_\tx c}$  \\ \midrule
       50      & 6133 & 48 & 85228 & 3.76  & 1.26   \\ \bottomrule
    \end{tabular}
\end{minipage}
\end{table}
The hyperelastic-plastic model describing the polymer matrix was fitted to experimental data of Fiedler et. al \cite{fiedler2001}, in which an epoxy matrix system is investigated for different loading cases. A comparison of the model response with the experimental results confirms a good agreement, see \reff{fig:fitting}. Only experimental tensile and compressive test data were used during the fitting process. Hence, the shear load case corresponds to a prediction and serves as a model validation. 

\begin{figure}
 \centering
 \linespread{1.3}
 \footnotesize
 \includegraphics{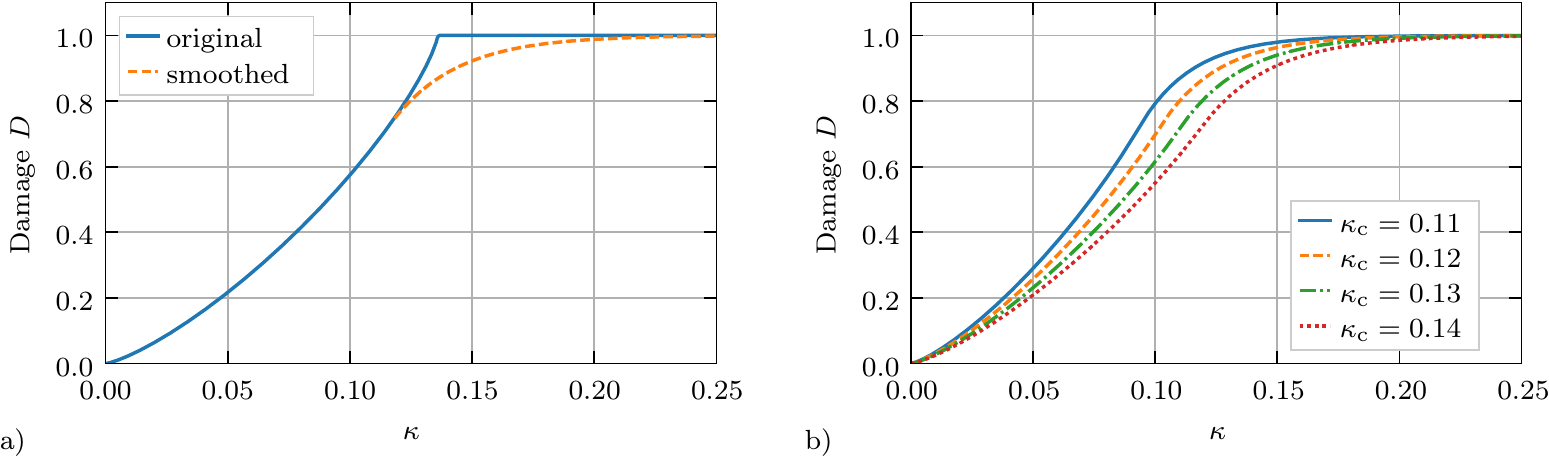}
 \caption{a) Damage evolution law in the original \refe{eq:damage_law} and smoothed \refe{eq:damage_law_mod} versions and b) damage evolution for different $\kappa_\tx c$ with fixed $b_1=b_2=0.3$ and $\kappa^\ast_\tx c=0.85\kappa_\tx c$.}
\label{fig:damage_law}
\end{figure}

The elastic constants as well as the plastic Poisson ratio in \reft{table:elastic_constants} are kept constant during the fitting procedure of the hardening laws and are taken from \cite{fiedler2001} and  \cite{melro2013}, respectively. Depending on the available stress-strain data, suitable hardening equations have to be defined in \refe{eq:harding_evo}. For the model response in \reff{fig:fitting}, a tensile hardening modulus of the type 
\begin{equation}
 h_\tx t(\eps^\tx p)=a_1 \exp(-a_2 \eps^\tx p)+a_3 (\eps^\tx p)^{a_4}
\end{equation}
has been used. Instead of proposing a similar ansatz for the compressive hardening modulus $h_\tx c$, a constant ratio $\sig_\tx t/\sig_\tx c$ between both hardening laws is assumed and identified as an additional fitting parameter to reduce the number of model parameters. The resulting hardening parameters are summarized in \reft{table:hardening}.
\\
The tensile and compressive strength for damage initiation in \refe{eq:damage_activation} can be read directly from the experimental results in \reff{fig:fitting}, due to the formulation of $F$ with respect to the Cauchy stress. However, the damage evolution, governed by \refe{eq:damage_evo} still needs a definition of the modulus $h_\tx D(D,\kappa)$. Since it is difficult to measure damage functions in general, simplified linear or exponential functions are commonly used. Various numerical tests during the model development indicate that linear functions often result in premature snapback events while exponential functions can lead to numerical difficulties, if the initial damage modulus is chosen too high. Therefore, the evolution law
\begin{equation}
\label{eq:def_h_D}
 h_\tx D\left( \kappa, D \right)= \frac{1}{\kappa_\tx c}\frac{b_1+1}{b_2+1} (1-D)^{-b_2} \left(\frac{\kappa}{\kappa_\tx c}\right)^{b_1}
\end{equation}
proposed in \cite{nguyen2019} is employed, which enables a flexible adjustment via the shape parameters $b_i$ and the critical damage driving parameter $\kappa_\tx c$. The definition in \refe{eq:def_h_D} seems to be a little extraordinary, but after integration a general polynomial damage law
\begin{equation}
\label{eq:damage_law}
 D=1-\left[ 1 - \left( \frac{\kappa}{\kappa_\tx c} \right)^{b_1+1} \right]^{\frac{1}{b_2+1}} 
\end{equation}
results. The damage law \refe{eq:damage_law} in its original form is shown in \reff{fig:damage_law}~a). While the parameters $b_1$ and $b_2$ adjust the shape of the curve, $\kappa_\tx c$ determines the point, where the damage variable $D$ reaches its maximum.  

If $D=1$, the material is fully damaged and should be excluded from the simulation. However, this point leads to additional numerical instabilities within an implicit solution procedure. As a suitable remedy the damage driving variable $\kappa$ in \refe{eq:damage_law} is replaced with
\begin{equation}
\label{eq:damage_law_mod}
\kappa^\ast=\left\{\begin{array}{ll} 
                    \kappa & \tx{for } \kappa \leq \kappa^\ast_\tx c \\
                    \kappa^\ast_\tx c + (\kappa_\tx c -\kappa^\ast_\tx c) \left(1-\exp\left[ - \frac{\kappa-\kappa^\ast_\tx c}{\kappa_\tx c-\kappa^\ast_\tx c} \right] \right) & \tx{for } \kappa > \kappa^\ast_\tx c
                    \end{array}\right.,
\end{equation}
leading to an exponential transition of $D$ towards $D=1$ after reaching the threshold $\kappa_\tx c$ \cite{nguyen2019}. The smoothed response is also shown in \reff{fig:damage_law}~a). The specific choice of $b_1=b_2$ and a fixed $\kappa^\ast_\tx c=0.85\kappa_\tx c$ are inspired by \cite{nguyen2019} and have been found appropriate regarding numerical stability. They are held fixed during the subsequent simulations, see \reft{table:damage}. The remaining parameters $\kappa_\tx c$ and the non-local length $l$ still influence the resulting failure process of a specimen under a specific loading direction. In this work, $l=0.45~r_\tx F$ is chosen to be smaller than the fiber radius in order to reduce the smoothing of a damaged zone over several fibers within an RVE. Finally, $\kappa_\tx c$ has to be determined in a way, that the model recovers the same critical energy release rate as observed in experiments. The sensitivity of the damage evolution with respect to $\kappa_\tx c$ is shown in \reff{fig:damage_law}~b). With increasing $\kappa_\tx c$ complete failure is delayed, leading to an increased energy dissipation. Typical values for the fracture energy $G_\tx c$ for epoxy polymers are in the range of $70-\SI{100}{\joule\per\square\metre}$ \cite{hsieh2010,chevalier2019}. According to \cite{nguyen2019}, $G_\tx c \approx \SI{85}{\joule\per\square\metre}$ is chosen and a virtual test procedure with a cylindrical bar was performed to determine a corresponding $\kappa_\tx c$. The parameters of the damage evolution are summarized in \reft{table:damage}. The detailed procedure of the virtual testing approach is described in \cite{nguyen2019}.
\begin{table}
\begin{minipage}[t]{0.45\textwidth}
    \centering
    \linespread{1.3}
    \caption{Parameters of the damage activation and evolution function for $G_\tx c = \SI{85}{\joule\per\square\metre}$\strut}
    \label{table:damage}
    \begin{tabular}{*{6}{c}}     \toprule
       $ \nicefrac{X_\tx t}{\si{\mega\pascal}}$& $\nicefrac{X_\tx c}{\si{\mega\pascal}}$ & $b_1$ & $b_2$ & $\kappa_\tx c$ & $\nicefrac{\kappa^\ast_\tx c}{\kappa_\tx c}$  \\ \midrule
       95      & 240 & 0.3 & 0.3 & 0.136  & 0.85   \\ \bottomrule
    \end{tabular}
\end{minipage}
\qquad\quad
\begin{minipage}[t]{0.45\textwidth}
    \centering
    \linespread{1.3}
    \caption{Parameters of the fiber-matrix interface\strut}
    \label{table:damage_interface}
    \begin{tabular}{*{4}{c}}     \toprule
       $\nicefrac{t^0_\tx n}{\si{\mega\pascal}}$&$\nicefrac{t^0_\tx{s}}{\si{\mega\pascal}}$ & $\nicefrac{G^\tx C_\tx{n}}{\si{\joule\per\square\metre}}$ & $\nicefrac{G^\tx C_\tx{s}}{\si{\joule\per\square\metre}}$ \\ \midrule
       50      & 70 & 50 & 70  \\ \bottomrule
    \end{tabular}
\end{minipage}
\end{table}
Finally, the properties of the interfaces have to be defined. The strength values $t^0_\tx n$ and $t^0_{\tx s}=t^0_{\tx s_{1}}=t^0_{\tx s_{2}}$ are taken from \cite{melro2013a}, where the choice is motivated by the investigations of \cite{vaughan2011}, see \reft{table:damage_interface}. Literature values for the interface fracture energy differ in orders of magnitude, depending on the fiber-matrix combination and the surface treatment of the fibers: values in the single-digit range \cite{melro2013a} over values around $\SI{100}{\joule\per\square\metre}$ \cite{wu2013}  up to several hundred $\si{\joule\per\square\metre}$ \cite{goldmann2018} are reported. In this study, a mean value is chosen, where the ratio between normal and shear fracture energy corresponds to the ratio between normal and shear strengths. All fiber-matrix parameters are summarized in \reft{table:damage_interface}.  
\\
Only the interface between aluminum and the epoxy matrix still needs a parameter set. It is difficult to determine suitable cohesive zone parameters of an interface along a microscopically rough surface. Hence, an initial simplified set, i.e. $t^0_\tx n=t^0_\tx s=\SI{15}{\mega\pascal}$, $G^\tx C_\tx{n}=G^\tx C_\tx{s}=\SI{20}{\joule\per\square\metre}$ , is chosen well below the strengths of the remaining fiber-matrix interfaces and the epoxy matrix. The influence of this parameter set will be studied additionally. In this study, the strengths will be increased to $40,45$ and $ \SI{50}{\mega\pascal}$, i.e. the range of the fiber-matrix interface, where $G^\tx C$ will also be increased by the same ratio. Furthermore, the isotropy of the interface will be varied to an anisotropic case, since shear strength and fracture energies are often higher than their normal counterparts. Here, a set with identical ratio is used according to the fiber-matrix interface, e.g. $R_\tx{cz}=t^0_\tx n/t^0_\tx s=\SI{50}{\mega\pascal}/\SI{70}{\mega\pascal}\approx0.71$ and an arbitrary lower ratio $R_\tx{cz}=0.5$.

\section{Results}
\label{sec:results}
\begin{figure}
\centering
 \linespread{1.3}
 \footnotesize
 \includegraphics{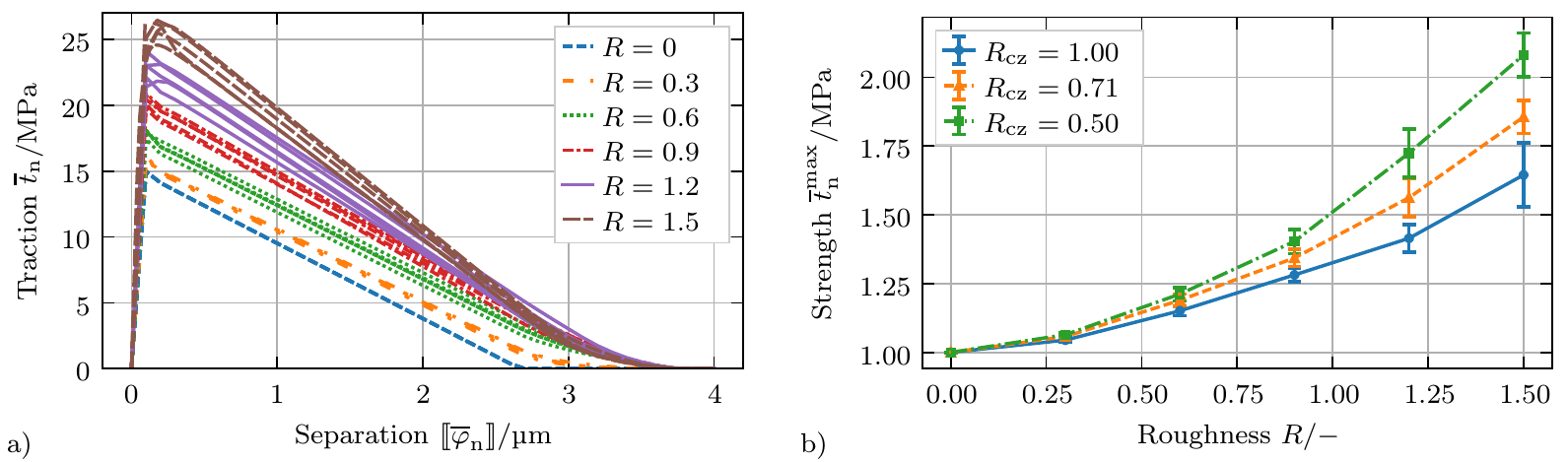}
 \caption{Results from homogenization of RVE with $t^0_\tx n=\SI{15}{\mega\pascal}$ and $G_\tx{n}^\tx C=\SI{20}{\joule\per\square\metre}$ under tension loading: a) Effective traction-separation law for $R_\tx{cz}=1.0$ and b) effective strength versus roughness $R$ for different cohesive ratios $R_\tx{cz}$. }
\label{fig:tsl_15MPa}
\end{figure}

In the following study, effective traction-separation laws are extracted by homogenization of several interface RVEs under tensile loading. The main objective is to investigate the relation of effective interface strength and the roughness of the local metal-polymer interface. The effective strengths are the maximum values of the effective tractions, which are calculated according to                                 \refe{eq:effective_tractions}. RVEs for each roughness $R \in \left\{0.0, 0.3, 0.6, 0.9, 1.2, 1.5\right\}$ were generated with the methodology presented in Section \ref{sec:rve_generation}. For every roughness 5 realizations are analyzed in order to calculate mean values. The correlation length $l_\tx{ACF}=\SI{5}{\micro\metre}$ is  constant, while $R_\tx q$ varies. Fibers with a radius of $r_\tx F=l_\tx{ACF}$ were placed with a target fiber volume fraction of $\phi=0.5$. All materials are parametrized according to Section \ref{sec:parametrization} and used for the subsequent study.\\
For the RVE, there are different loading cases possible to study the failure of an equivalent macroscopic interface: the pure mode cases of tensile and shear loading and a mix of both. Since the shear load and mixed-mode case lead to numerically complex contact states at the metal-polymer interface, this study is restricted to the pure tensile load case. 
\\
The results of the RVE simulations with interface strengths of $t^0_\tx n=t^0_\tx s=\SI{15}{\mega\pascal}$ and $G^\tx C_\tx{n}=G^\tx C_\tx{s}=\SI{20}{\joule\per\square\metre}$ are shown in \reff{fig:tsl_15MPa}~a). Each line represents one RVE realization and every color summarizes one interface roughness defined by \refe{eq:roughness}. For a flat interface, i.e. $R=0$, the homogenized or effective TSL recovers the bilinear shape of the underlying cohesive zone model with identical strengths and negligible variations. With increasing roughness $R$, the TSL becomes increasingly nonlinear, especially in the transition from the undamaged linear to the softening response. The observed increase in effective strength, as shown in \reff{fig:tsl_15MPa}~b), is caused only by an increase of the microscopic contact surface and therefore by an increased crack path.

\begin{figure}
 \centering
 \linespread{1.3}
 \includegraphics{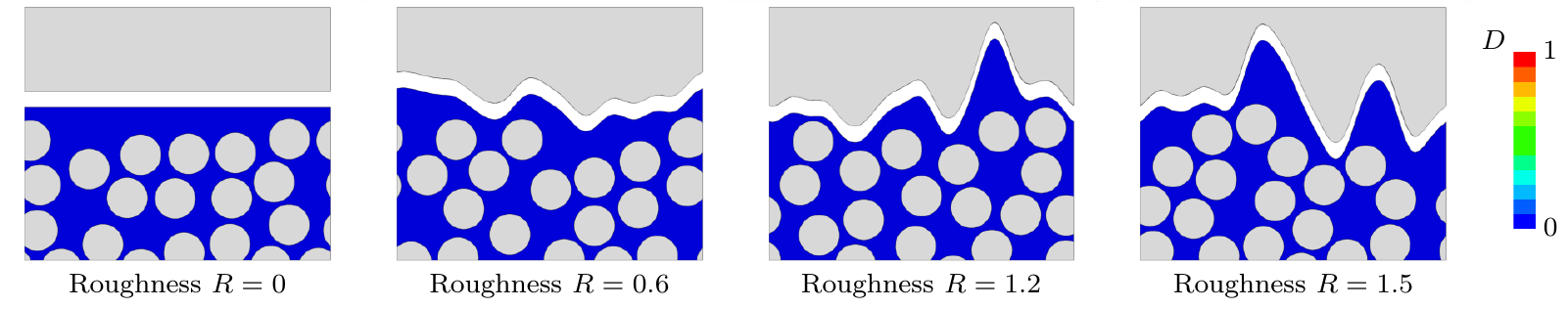}
 \caption{Contour plot of matrix damage variable $D$ in selected RVEs with $t_\tx n^0=\SI{15}{\mega\pascal}$, $G^\tx C_\tx{n}=\SI{20}{\joule\per\square\metre}$ and $R_\tx{cz}=1.0$. Only adhesive damage of the rough interface occurs due to the low strength of the interface.}
\label{fig:damage_plot_15MPa}
\end{figure}

For each roughness one RVE in the failed configuration is shown in \reff{fig:damage_plot_15MPa}. The only occurring damage mechanism is adhesive damage, since the interface strength is chosen far below any strength of the FRP material. The non-local damage variable $D$ of the matrix material would indicate cohesive polymer damage whereas the separation of the rough interface indicates adhesive damage. The influence of a certain anisotropy of the normal and shear parameters is also shown in \reff{fig:tsl_15MPa}~b). For this purpose, different values of the tangential strength $t^0_\tx s=t^0_\tx n/R_\tx{cz}$ and fracture energy $G^\tx C_\tx{s}=G^\tx C_\tx{n}/R_\tx{cz}$ computed from the same ratio $R_\tx{cz} \in \{1.0, 0.71, 0.5\}$ are analyzed.  $R_\tx{cz}=0.71$ corresponds to the constant ratio of cohesive properties in normal and tangential direction of the fiber-matrix interface, see \reft{table:damage_interface}. At low roughness values the local interface is primarily loaded in tensile mode. Hence, the effective interface strengths are almost the same. At high roughness values, the portion of the interface, which transfers shear stresses, increases. As a consequence, the effective interface strength is improved. 

\begin{figure}
\centering
 \linespread{1.3}
 \footnotesize
 \includegraphics{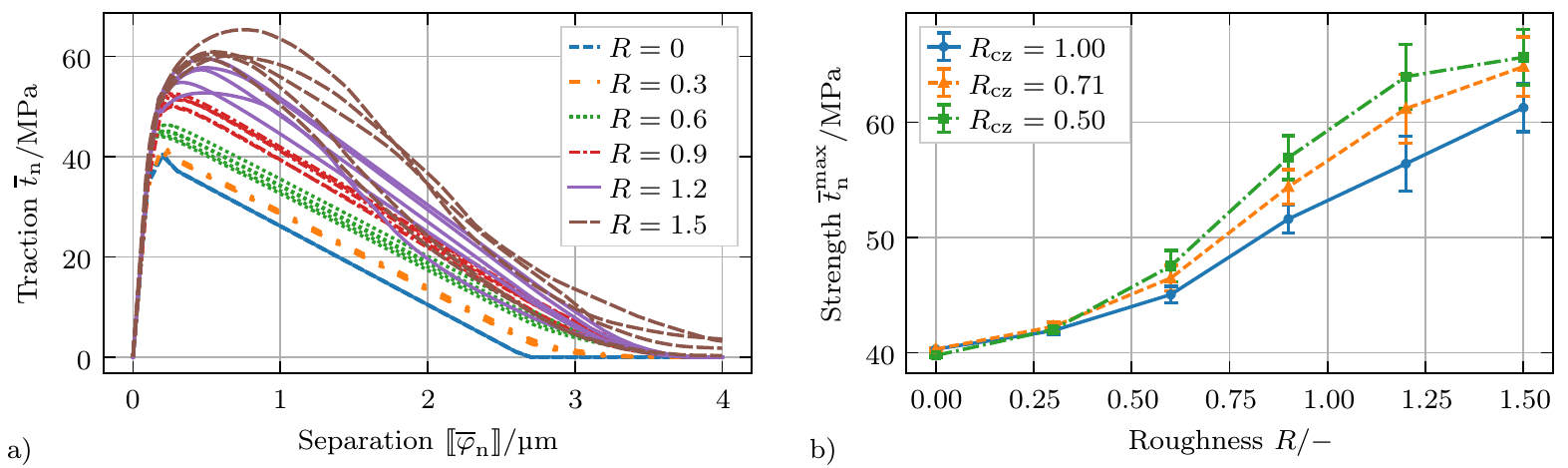}
 \caption{Results from homogenization of RVE with $t_\tx n^0=\SI{40}{\mega\pascal}$ and $G^\tx C_\tx{n}=\SI{53}{\joule\per\square\metre}$: a) Effective traction-separation law for $R_\tx{cz}=1.0$ and b) effective strength over roughness $R$ for different cohesive ratios $R_\tx{cz}$.}
\label{fig:tsl_40MPa}
\end{figure}

\begin{figure}[h]
 \centering
 \linespread{1.3}
 \includegraphics{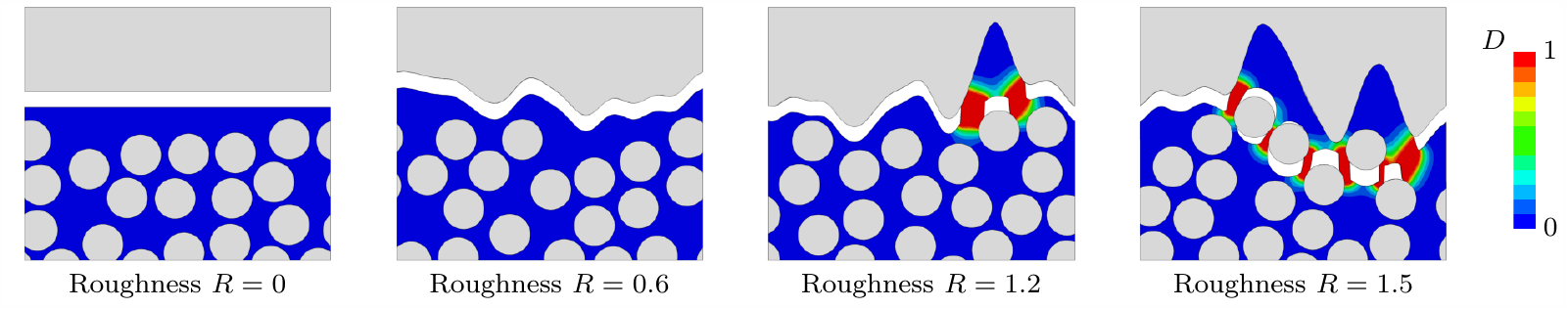}
 \caption{Contour plot of matrix damage variable $D$ in selected RVEs with $t_\tx n^0=\SI{40}{\mega\pascal}$, $G^\tx C_\tx{n}=\SI{53}{\joule\per\square\metre}$ and $R_\tx{cz}=1.0$.}
\label{fig:damage_plot_40MPa}
\end{figure}

\begin{figure}
 \centering
 \linespread{1.3}
 \includegraphics{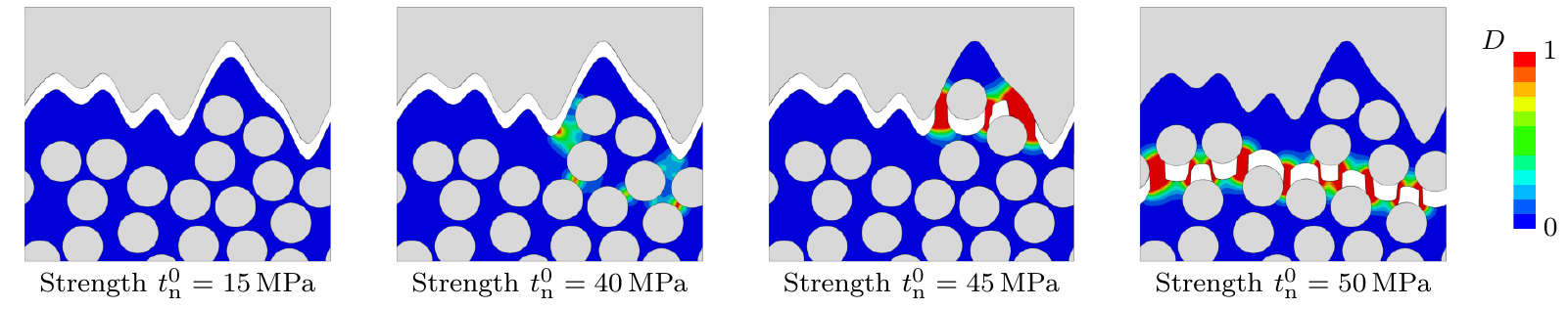}
 \caption{Contour plot of matrix damage variable $D$ in deformed configuration with $R=1.2$ for different interface strengths.}
\label{fig:damage_plot_CZ1p0}
\end{figure}

In the following investigation, the interface strength is increased to an arbitrary high value close to the strength of the FRP, i.e. the strength is set to $t_\tx n^0=\SI{40}{\mega\pascal}$ and the fracture energy is increased by the same ratio. The homogenized response is shown in \reff{fig:tsl_40MPa}. For a flat interface ($R=0$) the underlying TSL is still recovered, this time with higher strength and fracture energy. With increasing roughness, the response becomes strongly nonlinear, due to a change in the main failure mechanism. For an analysis, the contour plots of the non-local damage variable $D$ of the matrix material in \reff{fig:damage_plot_40MPa} are considered. Although the interface strength is increased, only adhesive failure occurs for low roughness values ($R=0$, $R=0.6$), which is indicated by a separation of the whole interface. For higher roughness values, also cohesive failure of the FRP material is observable. The main failure mechanism changes from purely adhesive failure of the metal-polymer interface to cohesive failure of the FRP. The influence of a tougher shear response of the metal-polymer interface is shown in \reff{fig:tsl_40MPa}~b). With low roughness values, the anisotropy is of minor influene and all effective strengths almost coincide. With increasing roughness, more shear components of the local TSL are coming into operation. This leads to an additional toughening effect and the shift from purely adhesive to cohesive failure occurs earlier. The curves in \reff{fig:tsl_40MPa}~b) show a trend to an absolute maximum value in the range of $\SI{70}{\mega\pascal}$, because the overall strength of the FRP acts as limiting component when the interface performance is fully exploited by the combination of high interface strength and roughness. This effect is shown in a last study and evaluated with the contour plots in \reff{fig:damage_plot_CZ1p0}. One failed realization of an RVE with an interface roughness of $R=1.2$ is shown for different local interface strengths $t^0_\tx n$. For simplicity, no anisotropy is used, i.e. $R_\tx{cz}=1.0$. Starting with purely adhesive failure at $t^0_\tx n=\SI{15}{MPa}$, 
an initiation of cohesive matrix damage is already observable for $t^0_\tx n=\SI{40}{MPa}$. Using $t^0_\tx n=\SI{45}{MPa}$, a mix of adhesive and cohesive failure occurs, where the global failure still localizes at the rough interface. With the highest investigated strength $t^0_\tx n=\SI{50}{MPa}$, the interface is no longer the weakest part of the material combination and the FRP always fails. A further increase of roughness or interface strength has no influence in this state, because the failure already localizes within the bulk material in a certain distance to the interface.

Lap joint tests or tests with a general mode II loading are commonly used in the literature to experimentally investigate the influence of surface roughness on the adhesive properties of metal-composite hybrids \cite{kim2010,lucchetta2011,zinn2018}. Experiments under tensile loading were carried out for simple material combinations in \cite{yao2002,cordisco2016,hosseini2019}, which confirm the qualitative results of the numerical study in this work. In \cite{hosseini2019}, pure adhesive failure of an interface with a sinusoidal pattern was investigated under mode I loading. As the ratio between the amplitude and the period of the sinusoidal pattern increases, the maximum strength also increases due to the larger contact area. This scenario corresponds to the case in \reff{fig:tsl_15MPa}~b) for $R_\tx{cz} = 1$ with a similar trend. The fraction of adhesive and cohesive failure of an epoxy polymer connected to a random rough surface was investigated in \cite{yao2002}. They also observed that at low roughness values only adhesive failure occurs and that with increasing roughness there is a transition from adhesive to cohesive failure with polymer residues on the surface. Nevertheless, further experimental investigations need to be performed to quantitatively validate the effective response of the micromodel.

\section{Conclusions}
\label{sec:conclusion}
The interface zone of hybrid material combinations is highly complex, and their performance is of special importance to guarantee operational reliability during lifetime. In this contribution, the connection of a glass fiber reinforced epoxy composite and an aluminum component is addressed. Complex geometries at the microscale, i.e. randomly distributed fibers and a rough interface shape, as well as a strongly nonlinear material behavior during the failure process are considered. A full experimental investigation on this length scale proves to be difficult, so that numerical analyses are suitable means. In this context, a modeling strategy which facilitates an extraction of effective traction-separation laws by performing numerical simulations of representative volume elements of the interface zone is proposed. Local damage phenomena are described as a combination of discrete failure of the interfaces and diffuse failure of the bulk material. To this end, cohesive zone and large strain elastoplastic damage models with gradient enhancement are applied. Several studies with varying interface roughness are presented. The investigations reveal that the homogenized interface properties improve with increasing roughness. Furthermore, the dominating failure mechanism highly depends on the local interface strength. Using a low interface strength, purely adhesive failure is formed independently of the interface roughness. Only when the local interface strength is increased to values close to the remaining material strengths, a shift from purely adhesive to cohesive failure is observable. In this context, cohesive failure is intended, because it indicates a highly performant interface. Both numerically investigated effects -- increased local interface strength and a higher roughness -- can be achieved with a pre-treatment as, e.g., sandblasting, a laser structuring process or a chemical pre-treatment.
\\ 
In this contribution only the tensile loading case is considered. The shear loading case is a meaningful investigation for future research. However, this loading case is very complex due to the contact sliding of the metal-polymer interface in combination with failure phenomena.

\section*{Acknowledgments}
The present project is supported by the German Research Foundation (DFG) within the Priority Program (SPP) 1712, KA3309/4-2 and the GWK by providing computing time through the Center for Information Services and HPC (ZIH) at TU Dresden on the HRSK-II. This support is gratefully acknowledged.

{\linespread{1.1}
\bibliography{references}
}
\end{document}